# Optical charge state manipulation of lead-vacancy centers in diamond

Yiyang Chen[1], Yoshiyuki Miyamoto[2], Eiki Ota[1], Ryotaro Abe[1], Takashi Taniguchi[3], Shinobu Onoda[4], Mutsuko Hatano[1], Takayuki Iwasaki[1,*]

[1]Department of Electrical and Electronic Engineering, School of Engineering, Institute of Science Tokyo, Meguro, 152-8552 Tokyo, Japan

[2]Advanced Power Electronics Research Center, National Institute of Advanced Industrial Science and Technology, Tsukuba, 305-8568 Ibaraki, Japan

[3]Research Center for Materials Nanoarchitectonics, National Institute for Materials Science, Tsukuba, 305-0044 Ibaraki, Japan

[4]Takasaki Advanced Radiation Research Institute, National Institutes for Quantum Science and Technology, 1233 Watanuki, Takasaki, 370-1292 Gunma, Japan

[*]Email: iwasaki.t.c5b4@m.isct.ac.jp

**Abstract**

Group-IV vacancy centers in diamond exhibit excellent optical and spin coherence properties, making them highly promising and scalable spin qubit candidates. Since only specific charge states are magneto-optically active, control over the charge state is fundamental for quantum applications. Here, we realize the charge state control of lead-vacancy centers (PbV) through multi-color laser irradiation. We achieve tunable population manipulation of the negatively charged state from 0 to 89%, paving the way for spin control of the negatively charged PbV center. Furthermore, through analysis of charge state dynamics, we propose a charge cycle between the neutral and negatively charged states, indicating a possible pathway to the neutral PbV center with a spin-1 system.

**Introduction**

Quantum emitters in diamond have emerged as one of the most leading platforms for quantum applications. Multi-node quantum networks have been demonstrated based on nitrogen-vacancy (NV) centers [1]. However, the fraction of zero-phonon line (ZPL) becomes as low as ~3%, leading to a limited entanglement generation rate [2]. To overcome this issue, the negatively charged group-IV color centers in diamond [3–6], including silicon-vacancy (SiV), germanium-vacancy (GeV), tin-vacancy (SnV), and lead-vacancy centers (PbV) with an inversion symmetry, have been proposed. These color centers exhibit higher Debye-Waller factors of 34-87% [7–11]. Long distance entanglement has been generated between SiV centers [12], while they require cooling to the mK range for a long spin coherence time [13] without a large strain [14]. The negatively charged PbV center possesses the highest zero filed splitting of ~3900 GHz in the ground state, effectively suppressing the phonon interaction between the sublevels. Therefore, the transform-limited linewidth has been observed even at temperatures above 10 K [11]. A spin coherence time on the millisecond level can be also expected at similar temperatures [6]. Accordingly, the PbV center is an important building block for quantum network nodes. In addition to the negatively charged state, the neutral state of the group-IV color centers is an interesting quantum system owing to the spin-1 system, leading to a long spin coherence time at an even higher temperature, as demonstrated in the SiV center [15].

Since only specific charge states are magneto-optically usable, understanding the charge transfer process and creating sufficient population in the desired charge states are fundamental for quantum application. Several approaches have been demonstrated for the charge state control of the color centers in diamond, including surface termination of diamond [16–18], doping with phosphorus [19] or boron [15], electrical tuning in devices [20–23], photo-carrier generation [24–26], and optical control [27–32]. Surface termination can be only applicable to color centers close to the surface, and dopants in the lattice possibly degrade the optical properties [15,19], which significantly limits further practical applications. Among the methods above, the optical irradiation is the simplest and highly controllable method, in which the coherent emission can be expected.

For the PbV center, capture of photo-carriers generated from optically-excited defects have been shown to lead to the charge state transition under 532 nm irradiation [33]. As the direct optical method, it has been known that 532 nm laser irradiation initializes PbV centers to the bright negatively charged state after their transition to a dark state under resonant excitation [11]. However, the important information on the transition rate, population, and transition mechanism are totally missing, and it has not been revealed whether the charge state can be optically cycled with non-resonant visible lasers. In this work, we report the charge state cycle of PbV centers by multi-color non-resonant laser irradiation. We demonstrate that blue laser irradiation shelves the PbV centers into a pure dark state with a one-photon process, while the green laser irradiation repump the charge state into the negatively charged state with a two-photon process. By combination of both lasers, we achieve directly and freely manipulating the charge state and population of the negatively charged state between around 0 to 89%, indicating that the green laser creates a sufficiently high population in the negatively charged state, towards high-fidelity spin control [34]. Finally, a model of the charge dynamics is proposed with the first-principles calculation.

**Results**
**Optical charge state control**

Figure 1(a) shows an atomic structure of the PbV center. A lead impurity takes an interstitial position between carbon vacancies, possessing the inversion symmetry with a nearly zero first-order permanent electrical dipole moment [35,36]. Under the combined influence of the spin-orbit interaction and Jahn-Teller effect [37,38], both ground and excited states of the negatively charged PbV center split into two sets of Kramers' doublets (Fig. 1 (b)). Consequently, four optical transition channels exist under zero magnetic field, mentioned as A-D. The C and D transitions have wavelengths of 550 nm and 554 nm, respectively, at a low temperature [6]. The frequency difference between the C and D transitions gives a large zero-field splitting of approximately 3900 GHz in the ground state. The C transition shows a transform-limited linewidth of ~39 MHz, while the D transition is significantly broadened due to the phonon interaction [11]. Thus, we measure the C transition of PbV emitters under the resonant excitation.

We investigate the charge state transition of a PbV center in Sample I (see Methods) under non-resonant laser irradiation (445 and 532 nm) using a sequence depicted in Fig. 1(c). After the first photoluminescence excitation (PLE) is recorded with the resonant laser, a 445 nm laser (28.5 µW) is irradiated. Then, the second PLE is sequentially measured. Following another 532 nm laser irradiation (100 µW), we record the third PLE spectrum. To avoid simultaneous irradiation of the resonant and non-resonant lasers, which may cause additional charge state transition as observed in GeV and SnV centers [30,31], the non-resonant laser is irradiated when the detuning of the tunable laser exceeds at least 4 GHz. As shown in Fig. 1(d), the first PLE spectrum shows a linewidth of 38 MHz by a Lorentzian function fitting, which agrees well with the transform-limited linewidth of PbV centers [11], indicating the high-quality formation of the PbV center. Note that, in the following two PLE scans, the zero detuning is uniformly set to the center wavelength of this first scan. In the second PLE scan, we do not see any peak in the scan range from +4.5 to -7.2 GHz detuning. The measurement data around the zero detuning is shown in Fig. 1(e). Finally, in the third scan after the application of the 532 nm laser irradiation, we again observe a resonant peak with a similar linewidth of 39 MHz at the photon frequency (Fig. 1(f)). These observations indicate that the 445 nm laser irradiation shelves the negatively charged PbV center into a dark state through photoionization, leading to the disappearance of the PLE spectrum, while the 532 nm laser irradiation repumps the PbV center into the bright negatively charged state, resulting in the reappearance of the peak. Note that we do not observe the ZPLs of the PbV- center under 445 nm irradiation in a PL spectrum, further manifesting the dark state transition of the PbV center.

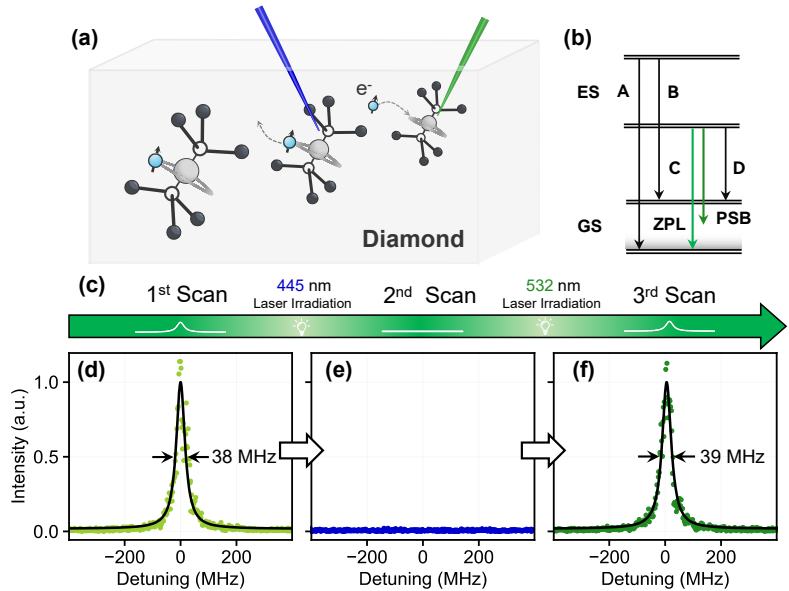

Figure 1. Charge state control of a PbV center using visible light laser irradiation. (a) PbV center in diamond. A blue laser changes the charge state of the PbV- center to another dark state, while a green laser initializes to the -1 charged state. (b) Optical transitions of PbV-. GS and ES denote the ground state and excited state, respectively. (c) A sequence employed to investigate the influence of non-resonant lasers to the charge state of PbV. A total of three PLE scans are continuously performed, with 445 nm and 532 nm laser irradiation at off-resonant frequencies (d) First PLE spectrum. (e) Second PLE spectrum after 445 nm laser irradiation. (f) Third PLE spectrum after 532 nm laser irradiation. The zero detuning of the three spectra is uniformly set to the central wavelength in the first scan in panel (d). The resonant laser power is set to 2 nW.

**Shelving process**

To reveal the dynamics behind the charge state transitions induced by the non-resonant lasers, we conduct time-resolved pulse experiments. The sequence in Fig. 2(a) is designed to investigate the charge state transition rate resulting from the 445 nm laser irradiation. 16 times repeated resonant and 15 times repeated non-resonant pulses correspond to the readout of the PbV charge state and charge state control, respectively. When the PbV center possesses the negatively charged state, it can be resonantly excited, producing a substantial number of photons in phonon sideband (PSB). On the other hand, the PbV center in the dark state cannot be resonantly excited, and thus, only background signals are detected. A subsequent 532 nm green pulse initializes to the negatively charged state. In Fig. 2(b), the fluorescence from the PbV$^-$ gradually decays as the number of the 445 nm pulse increases. Finally, it goes down to the background level, corresponding to a dark state. It is well known that the resonant excitation also results in the transition to the dark state. However, we find that the resonant laser power and the excitation time used here for the charge state read-out have little impact on this phenomenon. The fluorescence is recovered with a 532 nm laser pulse, indicating the charge state initialization into the bright negatively charged state. Figure 2(c) shows measured fluorescence as a function of the total irradiation time of the 445 nm laser at various 445 nm laser powers. At all laser powers, the fluorescence decays as increasing the irradiation time. The transition rates are obtained by fitting each curve with a mono-exponential function. The power dependence of the transition rate is presented in Fig. 2(d). A linear function fits well with a slope of 32 Hz/μW, indicating that the shelving process from the negatively charged state to the dark state is a one-photon process under the 445 nm laser irradiation.

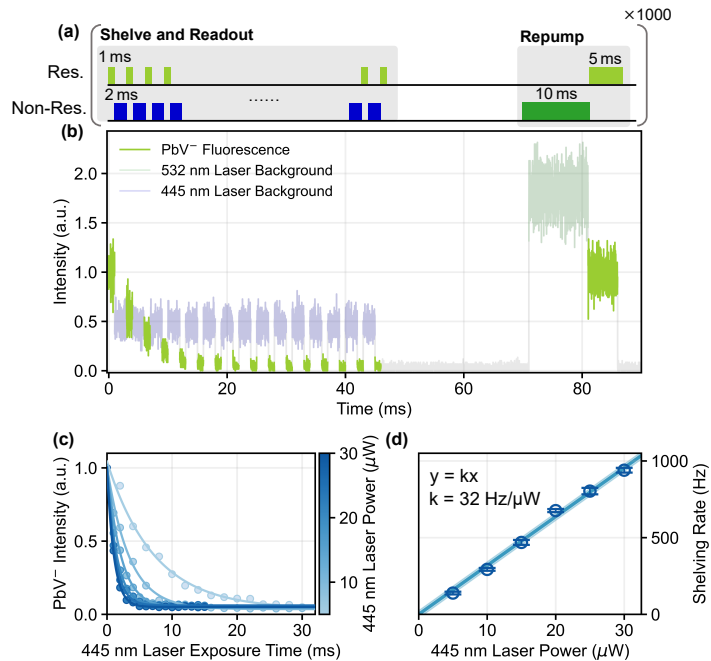

Figure 2. Time-resolved experiments of shelving process. (a) Pulse sequence. (b) Fluorescence from a negatively charged PbV center, controlled by 445 nm laser irradiation. (resonant laser: 2 nW, 445 nm laser: 10 μW, 532 nm laser: 100 μW) (c) Transition curves using different 445 nm laser powers.

(d) Transition rate as a function of the 445 nm laser power. Error bars are from the fitting error in panel (c) and the colored area represents the standard deviation of the linear fitting.

**Repump process**

Next, we perform the sequence of the repeated resonant readout and 532 nm charge control pulses (Fig. 3(a)). The 445 nm blue pulse resets the PbV center to the dark state, as seen in Fig. 2. In contrast to the result in Fig. 2(b), the fluorescence from the negatively charged state gradually recovers as increasing the 532 nm irradiation time (Fig. 3(b, c)). Interestingly, in Fig. 3(d), the repump rate has a non-linear behavior with a near-quadratic exponent of 1.81(0.26), suggesting that the repump mainly originates from a two-photon process. Aside from Sample I, we also perform the pulse sequence experiment shown in Fig. 3(a) on a color center in Sample II (see Methods) and observe a similar non-linear behavior. Among four randomly selected PbV centers from the two samples, three exhibit a similar non-linear behavior.

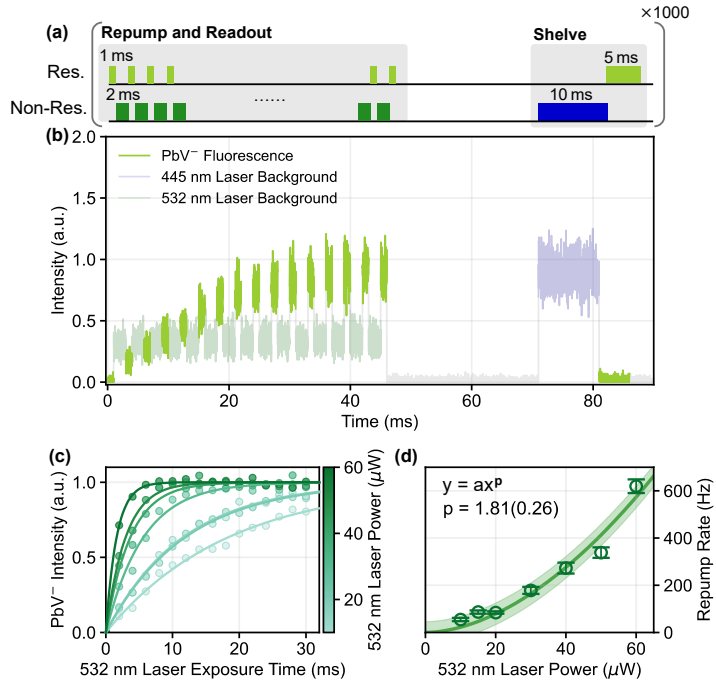

Figure 3. Time-resolved experiments of charge state initialization process. (a) Pulse sequence. (b) Fluorescence of a negatively charged PbV center, controlled by 532 nm laser irradiation. (resonant laser: 2 nW, 532 nm laser: 20 μW, 445 nm laser: 28.5 μW) (c) Transition curves using different 532 nm laser powers. (d) Transition rate as a function of the 532 nm laser power. Error bars are from the fitting error in panel (c) and the colored area represents the standard deviation of the power function fitting.

**Population**

Based on the pulse experiments above, we find that the fluorescence intensity of the PbV center can be freely controlled with the non-resonant lasers. This means that we can also manipulate the population in a certain charge state. Figure 4 shows the population analysis for a PbV by using the sequence in Fig.3 (a). Figure 4(a, b) show the distributions of the detected photons in 1 ms resonant readout repeated 1000 times with each 532 nm irradiation time and power. With the 532 nm irradiation (50 µW, 22 ms), a large peak centered at ~15 photon count appears, corresponding to the bright negatively charged state, while the low photon counts (<3) regime is attributed to the dark state (Fig. 4(a)). In contrast, without the 532 nm pulses after shelving with the 445 nm laser, we only obtain low counts in the histogram (Fig. 4(b)), indicating the formation of a pure dark state after the 445 nm laser irradiation. Figure 4(c) summarizes the population of the negatively charged state as functions of the 532 nm laser power and duration. Saturation behaviors are observed at high powers and long irradiation time domains, and the maximum population into the negatively charged state is ~89%.

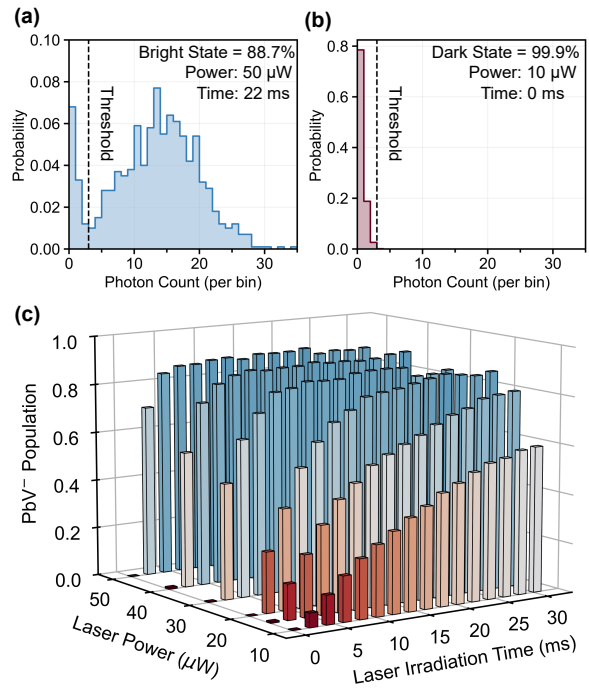

Figure 4. Manipulating the population of a PbV center. (a) The photon number distribution in a 1 ms charge state readout of a well-initialized case. (b) The photon number distribution after only 445 nm laser irradiation. (c) Population of the negatively charged state initialized at various 532 nm laser powers and times. The dashed lines in panels (a, b) are the threshold between the dark state and bright state that is set to three photons per bin.

**Mechanism of charge state transition**

Finally, we propose a model of the charge state transition of the PbV center under the non-resonant lasers. The calculated energy levels of the negatively charged and neutral states of the PbV center are demonstrated in Figure 5. First, we consider the shelving process caused by the 445 nm laser irradiation (Fig. 5(a)). The negatively charged PbV center has two possibilities for the photoionization: transition to the neutral state, or transition to the -2 negatively charged state. The transition to the neutral state by exciting an electron in the $e_g$ level to the conduction band of diamond requires an optical energy of ~2.6 eV [38], which is feasible with one 445 nm photon (2.79 eV). This agrees with the experimental observation in Fig. 2(d). On the other hand, a higher energy of ~3.5 eV is necessary for the transition to the -2 negatively charged state by capturing an electron from the valence band [38]. Additionally, the recovery from the -2 charged state requires only one green photon (~2.0 eV) [38], again contradicting the observation in Fig. 3(d). Thus, the dark state observed here is thought to be the neutral state.

The recovery process experimentally occurs in a two-photon process under the 532 nm laser irradiation (Fig. 3(d)). This agrees with the previous calculation [38] indicating that one 532 nm green photon (2.33 eV) does not have enough energy for the transition from the neutral to the -1 negatively charged state (~2.9 eV). Accordingly, as shown in Fig. 5(b), the first 532 nm photon is thought to excite the neutral PbV center into the excited state, creating a vacant in the $e_u$ level of the neutral state. Subsequently, the second 532 nm photon excites an electron from the valence band to occupy the vacant, consequently resulting in the charge state transition from the neutral state to the -1 negatively charged state. Thus, the neutral PbV initializes to the negatively charged state by two green photons.

It is worth noting that among the group-IV vacancy centers, the PbV center is only one emitter which can transition to the neutral state by visible blue laser irradiation according to the theoretical calculation [38]. A lighter group-IV emitter has a charge transition level between the negatively charged state and neutral state closer to the valence band of diamond in the energy gap, and all levels of SiV, GeV, and SnV centers are below the mid-gap. Indeed, the visible blue laser experimentally repumps the SiV, GeV and SnV center from the dark state into the negatively charged state [29,31,39]. In contrast, the energy levels of the PbV center shift upwards over the mid-gap, allowing the 445 nm laser to transition to the neutral state. Thus, the observed charge state transition is thought to be a unique process of the PbV center under 445 nm irradiation.

Finally, to investigate potential fluorescence from the neutral state, we examine PL spectra under the non-resonant lasers. The blue laser leads to observation of only one small peak at 715 nm. This emission frequently appears in previous reports in Pb implanted diamonds [6,40,41]. However, the emission wavelength is significantly different from predicted wavelengths of ~560 nm for the neutral PbV center in theoretical calculations [42,43]. Thus, we do not reach the conclusion about the origin of the observed 715 nm peak, and identification of the neutral charge state requires further investigations.

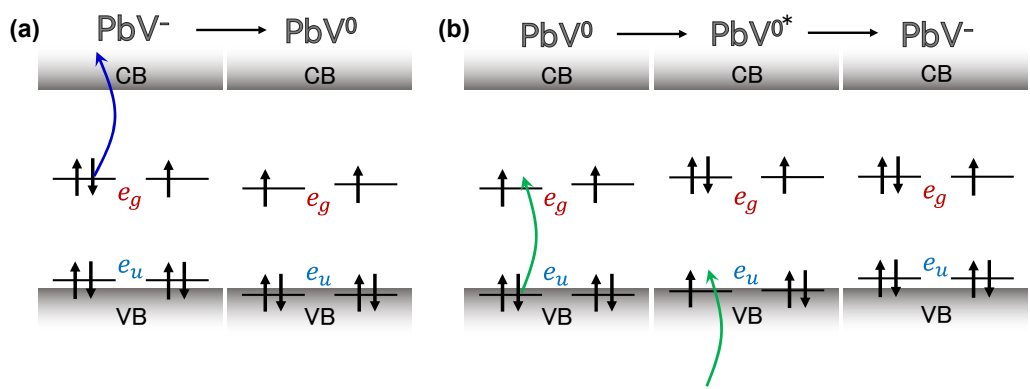

Figure 5. Proposed charge cycle process of the PbV center driven by (a) 445 nm and (b) 532 nm laser irradiation. VB and CB denote valence band and conduction band, respectively.

**Discussion**

The charge cycle model we find here advances the understanding of the charge state of the PbV center in the diamond host material. Under 532 nm laser irradiation, we obtain a high population up to 89% in the negatively charged state. This stable presence of the negatively charged state paves the way for the spin control of the PbV center, and establishes a solid foundation for spin single shot readout [34]. It is worth noting that the charge dynamics is likely to be affected by the laser wavelength. Thus, the excitation wavelength dependence [27] will be particularly interesting to fully understand the charge dynamics and stability of the PbV center. For the neutral state, we currently have no decisive evidence to determine whether the observed peak at 715 nm comes from the neutral PbV centers. Observation of the Stark effect of this emission line under external electric-fields could provide an insight into its atomic symmetry [35]. It will be also important to investigate the predicted wavelength (~560 nm). Since we do not see a peak at this wavelength under 445 nm laser irradiation, the addition of the resonant laser will enhance an excitation efficiency of this predicted line. Furthermore, the fabrication of a high-density PbV bulk sample could allow us to verify the existence of a spin-1 system in electron spin resonance measurements, as shown for the SiV center [15]. Lastly, we mention that remote charge state conversion by capturing photo-carriers occurs under 445 nm laser irradiation. Thus, we do not rule out the possibility that this process also plays a role in the charge state conversion observed above. However, direct laser irradiation should affect more efficiently, supported by the observation of a bright singularity spot for the charge state transition of NV, SiV, and PbV centers [25,26,33,44].

**Experimental and computational methods**

For the fabrication of the PbV centers, Pb ions are implanted into two IIa-type (001) single-crystal diamond substrates with a fluence of $5\times10^8$ cm$^{-2}$ at an acceleration energy of 12 MeV [33]. To restore the lattice damage induced by the ion implantation and to form the PbV centers, high pressure and high temperature (HPHT) annealing is performed at 2300°C (Sample I) or 2200°C (Sample II) under 7.7 GPa for 20 min [33].

All optical experiments are conducted in a home-built confocal microscope. The samples are cooled down to about 6 K by a closed-cycle helium cryostat (s50, Montana Instrument). The laser light is guided to the sample by an apochromat ×50 objective lens (MPLAPON50× NA=0.95, OLYMPUS). Blue (06-MLD 445 nm, Cobolt) and green (06-DPL 532 nm, Cobolt) non-resonant lasers are used to control the charge state of the PbV centers. Optical pulses of these lasers are generated by an acousto-optic modulator (AOM, 532 nm, rise time 25 ns, Gooch & Housego) or direct modulation equipped with the laser (445 nm, rise time <2.5 ns). For resonant excitation, a dye tunable laser (Matisse 2 DS, Sirah Lasertechnik) and a tunable diode laser (DL-SHG pro, Toptica) are employed. A wavelength meter (WS8-30, HighFinesse) with a resolution of 1 MHz is used to monitor the wavelength and stabilize the tunable lasers through PID control. To stabilize the power of the dye laser into a range of 2 - 20 nW, the noise eater (NEL01A/M, Thorlabs) is used throughout the experiment. The resonant laser is gated using another AOM (532 nm, rise time 25 ns, Gooch & Housego). Phonon-side band (PSB) is detected through an optical filter for the resonant excitation. To control the timing and duration of the laser pulses, an arbitrary waveform generator (Pulse Streamer, Swabian Instruments) is employed. For all time-resolved experiments, we count the electrical pulse signals from an avalanche photo-diode (SPCM-AQRH-14, Excelitas) using a high-resolution fast counter (Time Tagger Ultra, Swabian Instruments). The confocal mapping and PLE scan are recorded by the Qudi framework [45].

The first-principles calculations are performed within the density functional theory. The plane-wave basis set with cutoff energy of 64 Ry is used to express valence orbitals. Norm-conserving pseudopotentials [46] are used to express electron ion interactions. The pseudopotentials for Pb are constructed to treat Pb 5d orbitals as valence orbitals. The local density approximation using the functional form [47] is employed to express the exchange-correlation potentials. A supercell of 3×3×3 cubic diamond (216 C atoms) is used and the geometry optimization under electronic ground state is carried out by using the total-energy and force formalism [48]. All calculations are performed within the spin-unpolarized approximation with use of fractional occupation numbers.

**Acknowledgements**

This work is supported by JSPS KAKENHI Grant Number JP22H04962, the MEXT Quantum Leap Flagship Program (MEXT Q-LEAP) Grant Number JPMXS0118067395, JST Moonshot R&D Grant Number JPMJMS2062, and Council for Science, Technology and Innovation (CSTI), 3rd Cross-ministerial Strategic Innovation Promotion Program (SIP) Quantum.

**Reference**

[1]  M. Pompili et al., Realization of a multinode quantum network of remote solid-state qubits, Science (1979) **372**, 259 (2021).

[2]  H. Bernien et al., Heralded entanglement between solid-state qubits separated by three metres, Nature **497**, 86 (2013).

[3]  L. J. Rogers et al., Electronic structure of the negatively charged silicon-vacancy center in diamond, Phys. Rev. B **89**, 235101 (2014).

[4]  T. Iwasaki et al., Germanium-Vacancy Single Color Centers in Diamond, Sci Rep **5**, 12882 (2015).

[5]  T. Iwasaki, Y. Miyamoto, T. Taniguchi, P. Siyushev, M. H. Metsch, F. Jelezko, and M. Hatano, Tin-Vacancy Quantum Emitters in Diamond, Phys. Rev. Lett. **119**, 253601 (2017).

[6]  P. Wang, T. Taniguchi, Y. Miyamoto, M. Hatano, and T. Iwasaki, Low-Temperature Spectroscopic Investigation of Lead-Vacancy Centers in Diamond Fabricated by High-Pressure and High-Temperature Treatment, ACS Photonics **8**, 2947 (2021).

[7]  E. Neu, D. Steinmetz, J. Riedrich-Möller, S. Gsell, M. Fischer, M. Schreck, and C. Becher, Single photon emission from silicon-vacancy colour centres in chemical vapour deposition nano-diamonds on iridium, New J. Phys. **13**, 025012 (2011).

[8]  E. Neu, M. Fischer, S. Gsell, M. Schreck, and C. Becher, Fluorescence and polarization spectroscopy of single silicon vacancy centers in heteroepitaxial nanodiamonds on iridium, Phys. Rev. B **84**, 205211 (2011).

[9]  Y. N. Palyanov, I. N. Kupriyanov, Y. M. Borzdov, and N. V Surovtsev, Germanium: a new catalyst for diamond synthesis and a new optically active impurity in diamond, Sci Rep **5**, 14789 (2015).

[10] J. Görlitz et al., Spectroscopic investigations of negatively charged tin-vacancy centres in diamond, New J. Phys. **22**, 013048 (2020).

[11] P. Wang et al., Transform-Limited Photon Emission from a Lead-Vacancy Center in Diamond above 10 K, Phys. Rev. Lett. **132**, 073601 (2024).

[12] C. M. Knaut et al., Entanglement of nanophotonic quantum memory nodes in a telecom network, Nature **629**, 573 (2024).

[13] D. D. Sukachev, A. Sipahigil, C. T. Nguyen, M. K. Bhaskar, R. E. Evans, F. Jelezko, and M. D. Lukin, Silicon-Vacancy Spin Qubit in Diamond: A Quantum Memory Exceeding 10 ms with Single-Shot State Readout, Phys. Rev. Lett. **119**, 223602 (2017).

[14] P.-J. Stas et al., Robust multi-qubit quantum network node with integrated error detection, Science (1979) **378**, 557 (2022).

[15] B. C. Rose et al., Observation of an environmentally insensitive solid-state spin defect in diamond, Science (1979) **361**, 60 (2018).

[16] M. V Hauf et al., Chemical control of the charge state of nitrogen-vacancy centers in diamond, Phys. Rev. B **83**, 081304 (2011).

[17] Z.-H. Zhang et al., Neutral Silicon Vacancy Centers in Undoped Diamond via Surface Control, Phys. Rev. Lett. **130**, 166902 (2023).

[18] B. Grotz et al., Charge state manipulation of qubits in diamond, Nat Commun **3**, 729 (2012).

[19] J. Geng et al., Dopant-assisted stabilization of negatively charged single nitrogen-vacancy centers in phosphorus-doped diamond at low temperatures, npj Quantum Inf **9**, 110 (2023).

[20] M. Rieger, V. Villafañe, L. M. Todenhagen, S. Matthies, S. Appel, M. S. Brandt, K. Müller, and J. J. Finley, Fast optoelectronic charge state conversion of silicon vacancies in diamond, Sci Adv **10**,

eadl4265 (2024).

[21] T. Lühmann, J. Küpper, S. Dietel, R. Staacke, J. Meijer, and S. Pezzagna, Charge-State Tuning of Single SnV Centers in Diamond, ACS Photonics **7**, 3376 (2020).

[22] Y. Doi et al., Deterministic Electrical Charge-State Initialization of Single Nitrogen-Vacancy Center in Diamond, Phys. Rev. X **4**, 011057 (2014).

[23] C. Schreyvogel, V. Polyakov, R. Wunderlich, J. Meijer, and C. E. Nebel, Active charge state control of single NV centres in diamond by in-plane Al-Schottky junctions, Sci Rep **5**, 12160 (2015).

[24] Z.-H. Zhang, A. M. Edmonds, N. Palmer, M. L. Markham, and N. P. de Leon, Neutral Silicon-Vacancy Centers in Diamond via Photoactivated Itinerant Carriers, Phys. Rev. Appl. **19**, 034022 (2023).

[25] A. A. Wood, A. Lozovoi, R. M. Goldblatt, C. A. Meriles, and A. M. Martin, Wavelength dependence of nitrogen vacancy center charge cycling, Phys. Rev. B **109**, 134106 (2024).

[26] A. Wood, A. Lozovoi, Z.-H. Zhang, S. Sharma, G. I. López-Morales, H. Jayakumar, N. P. de Leon, and C. A. Meriles, Room-Temperature Photochromism of Silicon Vacancy Centers in CVD Diamond, Nano Lett. **23**, 1017 (2023).

[27] N. Aslam, G. Waldherr, P. Neumann, F. Jelezko, and J. Wrachtrup, Photo-induced ionization dynamics of the nitrogen vacancy defect in diamond investigated by single-shot charge state detection, New J. Phys. **15**, 013064 (2013).

[28] P. Siyushev, H. Pinto, M. Vörös, A. Gali, F. Jelezko, and J. Wrachtrup, Optically Controlled Switching of the Charge State of a Single Nitrogen-Vacancy Center in Diamond at Cryogenic Temperatures, Phys. Rev. Lett. **110**, 167402 (2013).

[29] J. Görlitz et al., Coherence of a charge stabilised tin-vacancy spin in diamond, npj Quantum Inf **8**, 45 (2022).

[30] K. Ikeda, Y. Chen, P. Wang, Y. Miyamoto, T. Taniguchi, S. Onoda, M. Hatano, and T. Iwasaki, Charge State Transition of Spectrally Stabilized Tin-Vacancy Centers in Diamond, ACS Photonics **12**, 2972 (2025).

[31] D. Chen, Z. Mu, Y. Zhou, J. E. Fröch, A. Rasmit, C. Diederichs, N. Zheludev, I. Aharonovich, and W. Gao, Optical Gating of Resonance Fluorescence from a Single Germanium Vacancy Color Center in Diamond, Phys. Rev. Lett. **123**, 033602 (2019).

[32] C. Pederson, N. S. Yama, L. Beale, M. Markham, M. E. Turiansky, and K.-M. C. Fu, Rapid, in Situ Neutralization of Nitrogen- and Silicon-Vacancy Centers in Diamond Using Above-Band Gap Optical Excitation, Nano Lett. **25**, 673 (2024).

[33] R. Abe, Y. Chen, P. Wang, T. Taniguchi, M. Miyakawa, S. Onoda, M. Hatano, and T. Iwasaki, Narrow Inhomogeneous Distribution and Charge State Stabilization of Lead-Vacancy Centers in Diamond, Adv. Funct. Mater. e12412 (2025).

[34] E. I. Rosenthal et al., Single-Shot Readout and Weak Measurement of a Tin-Vacancy Qubit in Diamond, Phys. Rev. X **14**, 041008 (2024).

[35] L. De Santis, M. E. Trusheim, K. C. Chen, and D. R. Englund, Investigation of the Stark Effect on a Centrosymmetric Quantum Emitter in Diamond, Phys. Rev. Lett. **127**, 147402 (2021).

[36] S. Aghaeimeibodi, D. Riedel, A. E. Rugar, C. Dory, and J. Vučković, Electrical Tuning of Tin-Vacancy Centers in Diamond, Phys. Rev. Appl. **15**, 064010 (2021).

[37] C. Hepp et al., Electronic Structure of the Silicon Vacancy Color Center in Diamond, Phys. Rev. Lett. **112**, 036405 (2014).

[38] G. Thiering and A. Gali, Ab Initio Magneto-Optical Spectrum of Group-IV Vacancy Color Centers

in Diamond, Phys. Rev. X **8**, 021063 (2018).

[39] J. A. Zuber, M. Li, Marcel. li Grimau Puigibert, J. Happacher, P. Reiser, B. J. Shields, and P. Maletinsky, Shallow Silicon Vacancy Centers with Lifetime-Limited Optical Linewidths in Diamond Nanostructures, Nano Lett. **23**, 10901 (2023).

[40] M. E. Trusheim et al., Lead-related quantum emitters in diamond, Phys. Rev. B **99**, 075430 (2019).

[41] S. Ditalia Tchernij et al., Spectral features of Pb-related color centers in diamond – a systematic photoluminescence characterization, New J. Phys. **23**, 063032 (2021).

[42] C. J. Ciccarino, J. Flick, I. B. Harris, M. E. Trusheim, D. R. Englund, and P. Narang, Strong spin–orbit quenching via the product Jahn–Teller effect in neutral group IV qubits in diamond, npj Quantum Mater **5**, 75 (2020).

[43] G. Thiering and A. Gali, The (eg ⊗ eu) ⊗ Eg product Jahn–Teller effect in the neutral group-IV vacancy quantum bits in diamond, npj Comput Mater **5**, 18 (2019).

[44] G. Garcia-Arellano, G. I. López-Morales, N. B. Manson, J. Flick, A. A. Wood, and C. A. Meriles, Photo-Induced Charge State Dynamics of the Neutral and Negatively Charged Silicon Vacancy Centers in Room-Temperature Diamond, Advanced Science **11**, 2308814 (2024).

[45] J. M. Binder et al., Qudi: A modular python suite for experiment control and data processing, SoftwareX **6**, 85 (2017).

[46] N. Troullier and J. L. Martins, Efficient pseudopotentials for plane-wave calculations, Phys. Rev. B **43**, 1993 (1991).

[47] J. P. Perdew and A. Zunger, Self-interaction correction to density-functional approximations for many-electron systems, Phys. Rev. B **23**, 5048 (1981).

[48] J. Ihm, A. Zunger, and M. L. Cohen, Momentum-space formalism for the total energy of solids, Journal of Physics C: Solid State Physics **12**, 4409 (1979).